\documentclass[fleqn,usenatbib]{mnras}

\usepackage{newtxtext,newtxmath}
\usepackage{txfonts}

\usepackage[T1]{fontenc}

\DeclareRobustCommand{\VAN}[3]{#2}
\let\VANthebibliography\thebibliography
\def\thebibliography{\DeclareRobustCommand{\VAN}[3]{##3}\VANthebibliography}

\usepackage{graphicx}	
\usepackage{amsmath}	
\usepackage{amssymb}	
\usepackage{multirow}
\usepackage{adjustbox}
\usepackage{comment}
\usepackage{tablefootnote}
\usepackage{ulem}

\newcommand{\lephare}{\texttt{LePHARE}}
\newcommand{\ebv}{$E\left(B-V\right)$}

\newcommand{\orcid}[1]{\href{https://orcid.org/#1}{\includegraphics[height=11pt]{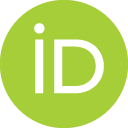}}}

\defcitealias{Jones2018}{J18}

\title[Photometric local SFR]{Reconsidering photometric estimation of local star formation environment and its correlation with Type Ia Supernova luminosity}

\author[Y.-L. Kim et al.]{
	Y.-L.~Kim$^{1, 2}$\thanks{E-mail: y.kim9@lancaster.ac.uk}\orcid{0000-0002-1031-0796}, 
	M.~Briday$^{2}$,
	Y.~Copin$^{2}$\orcid{0000-0002-5317-7518},	
	I.~Hook$^{1}$\orcid{0000-0002-2960-978X},
	M.~Rigault$^{2}$\orcid{0000-0002-8121-2560},
	M.~Smith$^{2}$\orcid{0000-0002-3321-1432}
\\
$^{1}$Department of Physics, Lancaster University, Lancs LA1 4YB, UK \\
$^{2}$Universit\'e de Lyon, Universit\'e Claude Bernard Lyon 1, CNRS/IN2P3, IP2I Lyon, F-69622, Villeurbanne, France \\
}

\date{Accepted 2023 November 12. Received 2023 November 12; in original form 2023 March 31}

\pubyear{2023}

\begin{document}
\label{firstpage}
\pagerange{\pageref{firstpage}--\pageref{lastpage}}
\maketitle

\begin{abstract}
Recent studies on the environmental dependence of Type Ia supernova (SN Ia) luminosity focus on the local environment where the SN exploded, considering that this is more directly linked to the SN progenitors.
However, there is a debate about the local environmental, specifically local star formation rate (SFR), dependence of the SN Ia luminosity.
A recent study claims that the dependence is insignificant ($0.051 \pm 0.020$ mag; $2.6\sigma$), based on the local SFR measurement by fitting local $ugrizy$ photometry data.
However, we find that this photometric local SFR measurement is inaccurate. 
We argue this based on the theoretical background of SFR measurement and the methodology used to make that claim with their local $ugrizy$ photometry data, especially due to a limited range of extinction parameters used when fitting the data.
Therefore, we re-analyse the same host galaxies with the same fitting code, but with more physically motivated extinction treatments and global $ugriz$ photometry of host galaxies.
We estimate global stellar mass and SFR.
Then, local star formation environments are inferred by using the method which showed that SNe Ia in globally passive galaxies have locally passive environments, while those in globally star-forming low-mass galaxies have locally star-forming environments.
We find that there is significant local environmental dependence of SN Ia luminosities: SNe Ia in locally star-forming environments are $0.072\pm0.021$ mag ($3.4\sigma$) fainter than those in locally passive environments, even though SN Ia luminosities have been further corrected by the BBC method that reduces the size of the dependence.
\end{abstract}





\section{Introduction}
\label{sec:intro}
Since the discovery of the accelerating universe from observations of distant Type Ia supernovae \citep[SNe Ia;][]{Riess1998, Perlmutter1999}, the environmental dependence of SN Ia luminosities is one of the major issues to understand the underlying physics of SNe and their luminosity evolution with redshift.
Over the past decades many studies found the trends between SN luminosity, after standard light-curve shape and color/extinction corrections, and global properties of host galaxies, such as stellar mass, star formation rate (SFR) and specific SFR (sSFR; the SFR per unit stellar mass), gas-phase metallicity, morphology, and stellar population age \citep[e.g.,][]{Neill2009, Kelly2010, Lampeitl2010, Sullivan2010, Childress2013, Pan2014, Campbell2016, Kang2016, Kim2019, Kang2020, Pruzhinskaya2020, Smith2020}.
Among those, the dependence of SN Ia luminosity on host stellar mass ($M_{stellar}$) is well-established: SNe Ia in low-mass hosts ($M_{stellar} \le 10^{10} M_{\sun}$) are 0.06--0.08 mag fainter than those in high-mass hosts.

Recent host studies focus on the local (i.e., kpc scale) environments at the SN explosion site, which are thought to be more directly linked to the SN progenitors.
Many studies found the same trend with similar size of the luminosity difference in a redshift range from 0.01 to 0.85: SNe Ia that exploded in locally star-forming and bluer in local $U-V\; (or\;R)$ color are $\sim$0.09 mag fainter than those in locally passive and redder environments \citep{Rigault2013, Rigault2015, Kim2018, Roman2018, Rigault2020, Kelsey2021}.
In particular, a recent study of \citet{Rigault2020} determined local sSFR as a tracer of a fraction of young stars at the SN location from H$\alpha$ emission measurements.
They estimated the H$\alpha$ emission within a projected radius of 1 kpc from the SN position by using the SuperNova Integral Field Spectrograph in the Nearby Supernova Factory \citep[SNfactory;][see details in \citealp{Rigault2013}]{Aldering2002, Lantz2004}.
From this, they found a luminosity difference of $0.163\pm0.029$ mag ($5.7\sigma$) between SNe Ia in locally star-forming and locally passive environments. 

However, \citet{Jones2015} and \citet[][here after \citetalias{Jones2018}]{Jones2018} argued that the dependence of SN Ia luminosities on the local sSFR is insignificant ($0.051 \pm 0.020$ mag; $2.6\sigma$). 
In particular, for the recent local sSFR study of \citetalias{Jones2018}, they derived local sSFRs by fitting local $ugrizy$ photometry within a circular 1.5~kpc radius aperture with the \lephare{}\footnote{\href{http://www.cfht.hawaii.edu/~arnouts/lephare.html}{http://www.cfht.hawaii.edu/$\sim$arnouts/lephare.h}} spectral energy distribution (SED) fitting code \citep{Arnouts1999, Ilbert2006, Arnouts2011}. 
As pointed out by \citetalias{Jones2018}, their photometric estimation of the local sSFR using $ugrizy$ data, however, is expected to be neither as accurate nor as precise as spectroscopic estimation, especially at the kpc scale \citep[see][]{Calzetti2013}.

This has been confirmed by \citet{Briday2022}. 
Based on a dataset of \citet{Rigault2020}, they derived various literature environmental tracers, such as global mass, local color and local photometric sSFR as in \citetalias{Jones2018}.
Then, they compared each tracer with spectroscopic local sSFR estimated by using the H$\alpha$ emission line, assuming that this measurement is a reference tracer.
Accounting for the measurement errors, they found that 37\% of photometric local sSFR  environments are misclassified, such that locally star-forming environments are classified as locally passive or conversely.

Yet, the local spectroscopic sSFR measurement from the H$\alpha$ emission is likely inaccessible for most SNe Ia surveys, especially at the high-redshift range. 
Given the impact of host galaxy properties on the inferred luminosity of SNe Ia and consequently on the derivation of cosmological parameters, such as $w$ or $H_0$ \citep[e.g.,][]{Rigault2015, Rigault2020}, a photometric estimation would be convenient.
We thus reconsider the accuracy and precision of photometric derivation of the local sSFR in that context.

In Sec.~\ref{sec:reconsider}, first, we review the theoretical basis describing the local scale SFR measurement through the SED fitting of photometric data.
Then, we investigate the local sSFR estimation method described in \citetalias{Jones2018}, where we highlight sensitivity of \lephare{} input parameters, especially extinction treatments, used by \citetalias{Jones2018} (Sec.~\ref{sec:impact}).
We discuss our result on the environmental dependence in Sec.~\ref{sec:result}, where we notably reconsider the method introduced by \citet{Kim2018}, which employs global host properties to infer local environments.
We conclude in Sec.~\ref{sec:discussion}.

\section{Photometric local star formation rate measurements}
\label{sec:reconsider}

\subsection{Theoretical basis of local SFR measurements through the SED fitting method}
\label{subsec:theory}

Since being introduced by \citet{Kennicutt1998}, H$\alpha$ emission is known as one of the most reliable and widely-used direct measurements for the galaxy absolute SFR \citep[see, e.g.,][]{Hopkins2003, Kennicutt2012, Calzetti2013}.
As explained in \citet{Calzetti2013}, this spectroscopic SFR estimation provides the number of stars formed within the last 100 Myr.

Photometric derivation of SFR, based on SED-fitting of \textit{global} photometric measurements, are calibrated against H$\alpha$-based SFR derivations \citep[see e.g.,][]{Salim2007, Wuyts2011, Smith2012, Childress2013}.
For example, \citet{Childress2013} showed that these global photometric SFR measurements based on broadband photometry (UV+Optical+IR) show good agreement with a typical $\sim$0.2 dex dispersion.
This scatter is thought to reflect that the SED fitting method measures the mean SFR averaged over longer time period ($\sim$500 Myr) than what H$\alpha$ emission probes \citep[see][]{Kennicutt2012, Calzetti2013}.

Measuring \textit{local} SFR is more complex and uncertain \citep{Kennicutt2012, Calzetti2013}.
\citet{Calzetti2013} indeed highlights that current SFR calibrations are only applicable to whole galaxies that are, in first approximation, evolved and isolated systems.
However, at local scale, it is unclear how the stellar initial mass function is sampled and whether the H$\alpha$ to SFR proportionality constants is unique, because local scale is not an isolated system.
\citet{Teklu2020} further remark the risk of using a single subset of attenuation curve for star-forming regions.

H$\alpha$-based spectroscopic local SFR, however, still directly probes the number of very young stars responsible for ionizing the circumstellar gas.
The shorter timescale measurements probed by spectroscopic SFR are less dependent on the evolution of the stellar population and \citet{Calzetti2013} suggests that ionizing photon tracers like H$\alpha$ may "fit the bill".
In the context of SN host property analysis, this supports the accuracy of H$\alpha$-based spectroscopic local sSFR introduced by \citet{Rigault2020}.
Concerning SED-based local SFR, recent works suggest that a panchromatic method, combining UV to submm wavelength ranges, may overcome most of the known problems  \citep[see e.g.,][]{Wright2016, Baes2020}.

Unfortunately, it is difficult to access panchromatic local information for targets in the redshift range covered by SNe Ia.
Hence, \citetalias{Jones2018} only used $ugrizy$ data to derive their local SFR.
Given the aforementioned theoretical limitations associated with SED-based local SFR estimation that uses only $ugrizy$ data and empirically shown by \citet{Briday2022} (see Sec.~\ref{sec:intro}), the accuracy of \citetalias{Jones2018} local SFR measurements is questionable.

\begin{table}
\centering
\caption{BC03 templates in \lephare{} and the extinction treatment used in \citetalias{Jones2018} and this work.}
\label{tab:bc03}
\begin{tabular}{c c c | c c c}
\hline\hline\\[-0.8em]
Template & Metallicity & $\tau$        && \multicolumn{2}{c}{Extinction applied} \\ \cline{5-6}
number   &  ($Z$)         & (Gyr)    &&  \citetalias{Jones2018} & This work \\
\hline \\[-0.9em]
1 & \multirow{9}{*}{0.004}  & 0.1 && - & \checkmark \\
2 &  & 0.3 && - & \checkmark \\
3 &  & 1  && - & \checkmark \\
4 &  & 2  && \checkmark & \checkmark \\
5 &  & 3  && \checkmark & \checkmark \\
6 &  & 4  && \checkmark & \checkmark \\
7 &  & 10 && \checkmark & \checkmark  \\
8 &  & 15 && \checkmark & \checkmark \\
9 &  & 30 && -  & \checkmark \\
\hline
10 & \multirow{9}{*}{0.008}  & 0.1 && - & \checkmark \\
11 &  & 0.3 && - & \checkmark \\
12 &  & 1  && - & \checkmark \\
13 &  & 2  && - & \checkmark \\
14 &  & 3  && - & \checkmark \\
15 &  & 4  && - & \checkmark \\
16 &  & 10 && - & \checkmark \\
17 &  & 15  && - & \checkmark \\
18 &  & 30  && - & \checkmark \\
\hline
19 & \multirow{9}{*}{0.02 ($Z_{\odot}$)}  & 0.1 && - & \checkmark \\
20 &  & 0.3 && - & \checkmark \\
21 &  & 1  && - & \checkmark \\
22 &  & 2  && - & \checkmark \\
23 &  & 3  && - & \checkmark \\
24 &  & 4  && - & \checkmark \\
25 &  & 10 && - & \checkmark \\
26 &  & 15  && - & \checkmark \\
27 &  & 30  && - & \checkmark \\
\hline
\end{tabular}
\end{table}

 \subsection{Method: Input parameters}
 \label{subsec:input}
In order to reproduce other results, it is critical to use the same data under the same conditions.
In the observational data analysis, these are the same photometry data and the same input parameters, respectively.
In particular, results from the SED fitting method are very sensitive to the input parameters, such as SED templates and an extinction treatment.
Because \citetalias{Jones2018} do not provide their local photometry data and sufficient information about the input parameters they used for \lephare{} (version 2.2), we directly asked D. Jones, and he kindly provided us the information that we required. 

In their configuration file, we find that \citetalias{Jones2018} employed one of the \lephare{} predefined SED libraries: \citet[][hereafter BC03]{BC2003} library.
In a BC03 README file in \lephare{}, this library contains 27 SED templates from three different metallicities and 9 different SFR $e$-folding timescales ($\tau$) with an exponentially decreasing star formation rate (Tab.~\ref{tab:bc03}).
In addition, it employs the Chabrier initial mass function \citep{Chabrier2003} and does not include extinction.
When we run \lephare{}, one then needs to define \textit{`$MOD\_EXTINC$'} parameter: the range of models for which extinction will be applied.
For the extinction treatment, \citetalias{Jones2018} used the Allen MW extinction law \citep{Allen1976} with \ebv{} values from 0 to 0.4 mag in a 0.1 step.
Then, \citetalias{Jones2018} applied the extinction law to the model numbers between 4 and 8 (see the Table~\ref{tab:bc03}), not for the entire (or most) of the 27 BC03 templates. 

The BC03 templates, which are employed by \citetalias{Jones2018}, are theoretical ones and do not include extinction.
BC03 templates do not correspond to a specific morphology, but differ by metallicities and SFR timescales.
Therefore, the extinction law should be applied to all (or most) of the 27 BC03 templates, instead of between model numbers 4 and 8.
From this finding, we first are curious about an impact of the  extinction treatment of \citetalias{Jones2018} on the \lephare{} results, especially on the SFR measurements, so we investigate this in the next section.

\begin{table*}
\centering
\caption{Differences in \lephare{} input parameters used in this work and \citetalias{Jones2018}.}
\label{tab:lephare_config}
\begin{tabular}{l c c c c c}
\hline\hline\\[-0.8em]
Parameters                            & \textit{MOD\_EXTINC}  &\textit{EB\_V} & \textit{EXTINC\_LAW} & \textit{ERR\_SCALE (ugriz)} & \textit{Z\_STEP}  \\
\hline \\[-0.9em]
\citetalias{Jones2018} & 4 8   & 0.0,0.1,0.2,0.3,0.4 &  \textit{MW\_Allen.dat} &  0.02,0.02,0.02,0.02,0.02  & 0.01,0.9,1 \\ [0.5em]
\multirow{3}{*}{This work}   &         & 0.0,0.05,0.1,0.15,0.2, \\
                                           & 1 27 & 0.25,0.3,0.35,0.4,0.45, &  \textit{LMC\_Fitzpatrick.dat} &  0.05,0.02,0.02,0.02,0.03 & 0.004,0.10,0.1 \\
                                            &        & 0.5,0.6,0.7,0.8,0.9,1.0 \\
\hline
\end{tabular}
\end{table*}

\section{Impact of the extinction treatment on \lephare{} results}
\label{sec:impact}

We run \lephare{}  by using its python wrapper \textit{\texttt{pylephare}}\footnote{\href{https://github.com/MartinBriday/pylephare}{https://github.com/MartinBriday/pylephare} }.
We employ the same BC03 templates as \citetalias{Jones2018} with our configuration and \citetalias{Jones2018} configuration to compare their results.
Differences in our input parameters from  \citetalias{Jones2018} are (Tab.~\ref{tab:lephare_config}): \\

\begin{enumerate}

\item[i.] \textit{`MOD\_EXTINC \hspace{0.1cm} 1 27'} \\
--we apply the extinction treatment to all the BC03 templates that do not include extinction (Tab.~\ref{tab:bc03}).

\item[ii.] \textit{`EB\_V \hspace{0.1cm}  0.0,0.05,0.1,0.15,0.2,0.25,0.3,0.35,0.4,0.45,0.5,0.6,
0.7,0.8,0.9,1.0'} \\
--\citet{Amanullah2015} and \citet{Johansson2021}, who estimated the host \ebv{} from SN light-curves, showed that the \ebv{} range is wider than the range in \citetalias{Jones2018} configuration (from 0.0 to 0.4).

\item[iii.] \textit{`EXTINC\_LAW \hspace{0.1cm} LMC\_Fitzpatrick.dat'} \\
--we use the \citet{Fitzpatrick1986} extinction law for Large Magellanic Cloud.
Using \citet{Allen1976} for the Milky Way has little effect on the \ebv{} distribution (see green and indigo histograms mostly overlapped in Fig.~\ref{fig:comp_ebv_all}).

\item[iv.] \textit{`ERR\_SCALE \hspace{0.1cm} 0.05,0.02,0.02,0.02,0.03'} \\
--we include SDSS's suggested error floor for $ugriz$ bands \citep[see \href{http://kcorrect.org}{kcorrect.org} and][]{Childress2013}.

\item[v.] \textit{`Z\_STEP \hspace{0.1cm} 0.004,0.10,0.1'} \\
--we use more grids to increase fit quality.

\end{enumerate}

\begin{figure}
	\centering
	\includegraphics[width=0.45\textwidth]{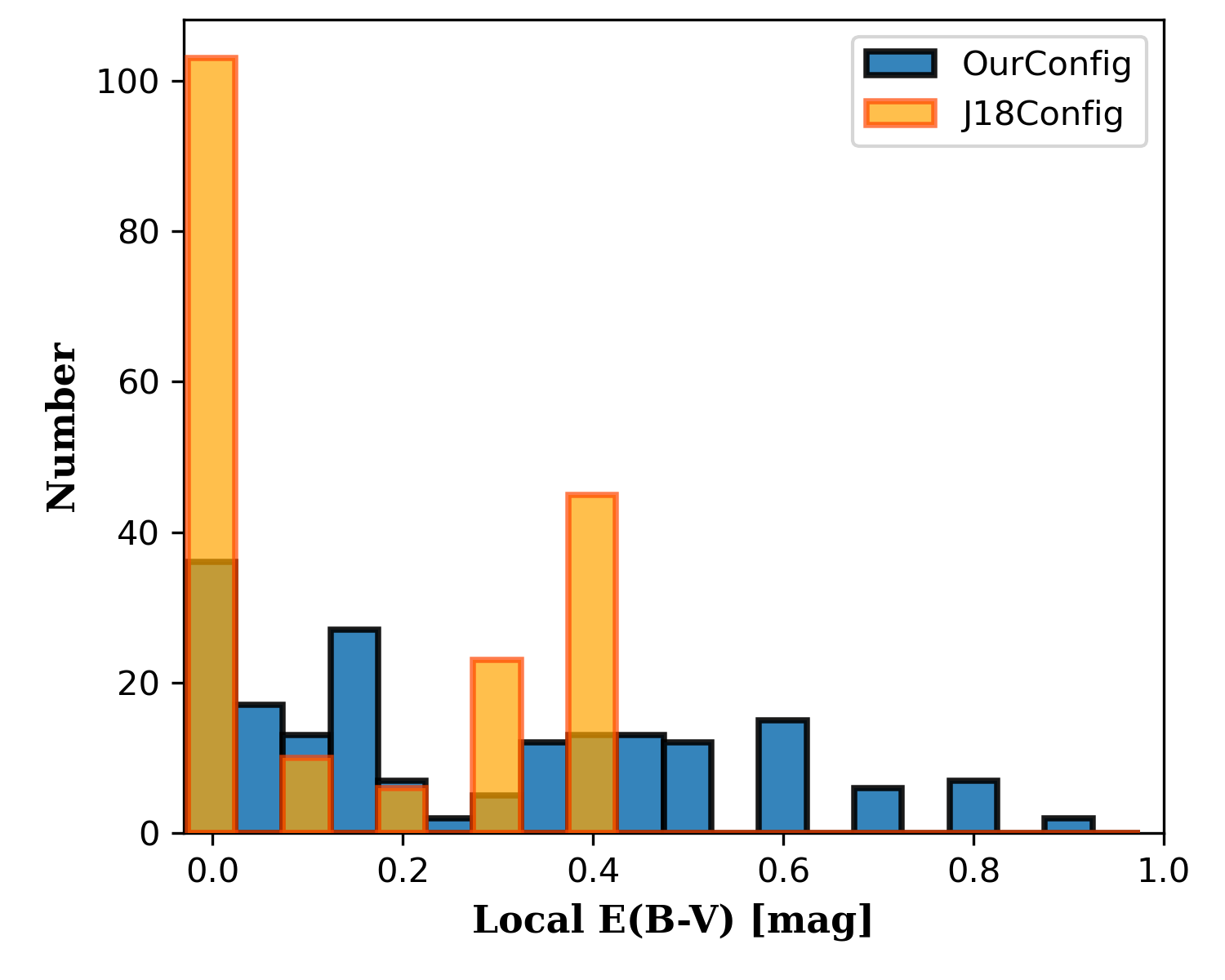}
	\includegraphics[width=0.45\textwidth]{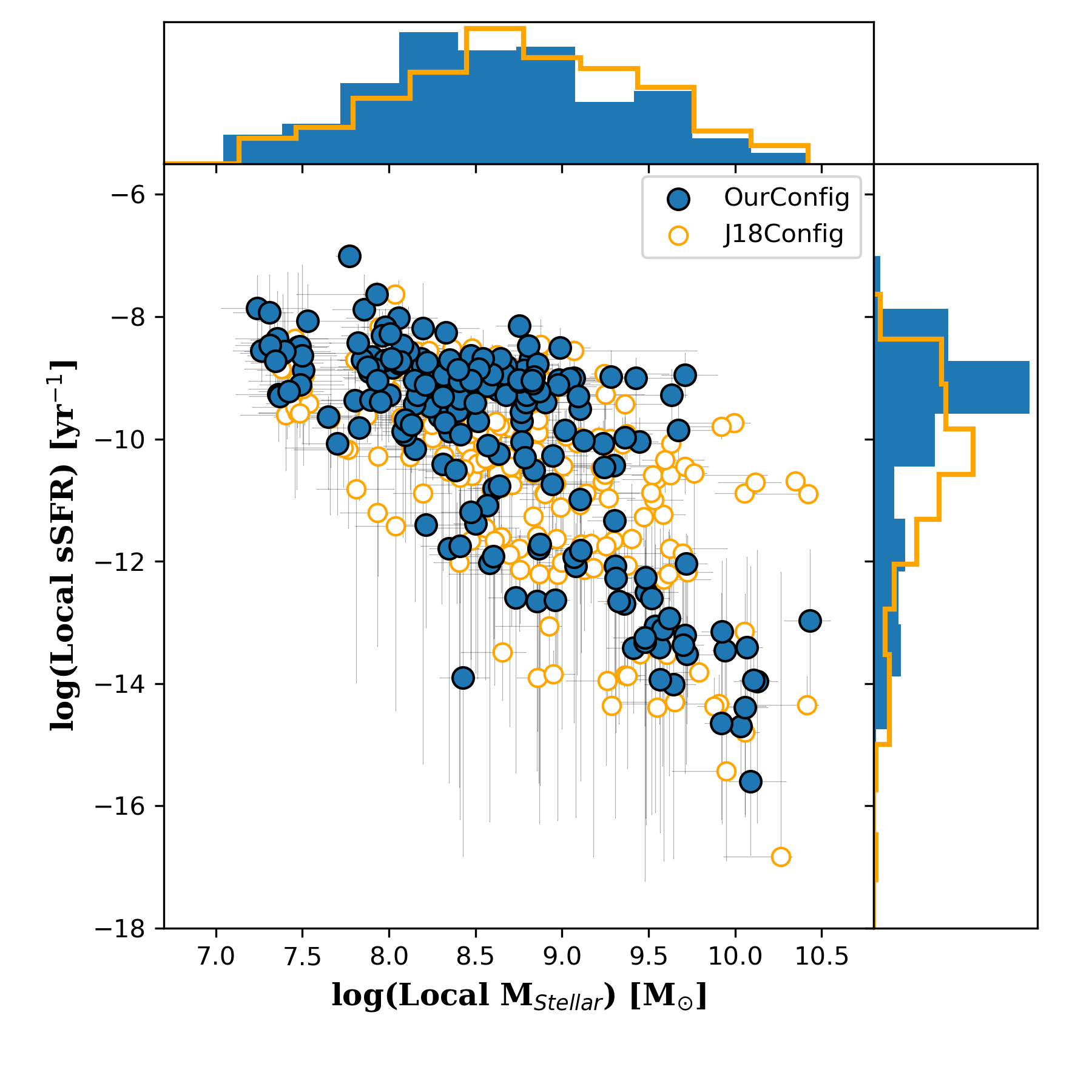}
	\caption{
	Distribution of local \ebv{} (top panel) and local $M_{stellar}$ versus local $sSFR$ (bottom) determined by local $ugrizy$ photometry data of \citetalias{Jones2018} and \lephare{} with our (blue) and \citetalias{Jones2018} (orange) configurations.
	Local environments with \ebv{} = 0.0 mag are dominant (55\%; 103 out of 187) when using the \citetalias{Jones2018} configuration.
	Local $sSFR$ measurements show a 1.0 dex (a factor of 10) difference between using our and \citetalias{Jones2018} configurations.
	}
	\label{fig:local_ebv_mass_ssfr}
\end{figure}

\subsection{Test with local photometry data}
\label{subsec:local}

First, we run \lephare{} with \citetalias{Jones2018}'s local $ugrizy$ photometry data we obtained directly from D. Jones.
Among 189 local environment data, two (2002ha and SN300105) failed to fit.
\lephare{} returned $\chi^{2}$ = 461.8 and 2583.5 for SN 2002ha local environment data and $\chi^{2}$ = 1006.6 and 1005.2 for SN 300105, for our and \citetalias{Jones2018} configurations, respectively.

Fig.~\ref{fig:local_ebv_mass_ssfr} shows a clear difference in the distributions of local \ebv{} and local $sSFR$ measurements between using our and  \citetalias{Jones2018} configurations.
The most notable point is that local environments with \ebv{} = 0.0 mag are dominant (55\%; 103 out of 187) when using the \citetalias{Jones2018} configuration (orange solid histogram).
In contrast, when using our configuration (blue solid histogram), all local samples are well-distributed over a wide range of \ebv{}.
The fraction of local environments with \ebv{} = 0.0 in our case is 19\% (36 out of 187).

The bottom panel of the figure presents local mass and local $sSFR$ distributions.
The local mass measurements show a similar distribution with a 0.1 dex difference, while the local $sSFR$ measurements differ by 1.0 dex (a factor of 10) between using our and \citetalias{Jones2018} configurations.
However, because the local $sSFR$ estimation from the SED fitting method is expected to be highly uncertain as we discussed in Sec.~\ref{subsec:theory}, we can not determine which one is correct to perform a more detailed analysis on the impact of the extinction treatment on the \lephare{} results.
Therefore, global photometry data is required.

\subsection{Test with global photometry data}
\label{subsec:global}

Global properties of galaxies, such as stellar mass and SFR, estimated from the SED fitting method are well established and show good agreement with those from spectroscopic data (see Sec.~\ref{subsec:theory}).
Regarding the extinction and reddening measurements, \citet{Meldorf2022} showed that the inclusion of near-infrared bands to optical data improve the precision of a dust parameter, but dose not significantly change the mean value (see their figure 2). 
Therefore, in order to test the impact of the extinction treatment on the global properties of galaxies, we use the SED fitting method with global optical photometry data.

We first obtained host galaxy $ugriz$ photometry data (`$modelMag$' and `$modelMagErr$') from the Sloan Digital Sky Survey 16th data release \citep[SDSS DR16;][]{Ahumada2020} with host galaxy coordinates provided by \citetalias{Jones2018}.
For the MW extinction correction, we also queried `$extinction$' magnitude for each filter.
Among 273 host galaxies in \citetalias{Jones2018}, 194 are in SDSS DR16.

Then, from the MW extinction-corrected $ugriz$ magnitude ($modelMag - extinction$), we run \lephare{}.

Our \lephare{} fit results are presented in Tab.~\ref{tab:data}.

\begin{figure}
	\centering	
	\includegraphics[width=1\columnwidth]{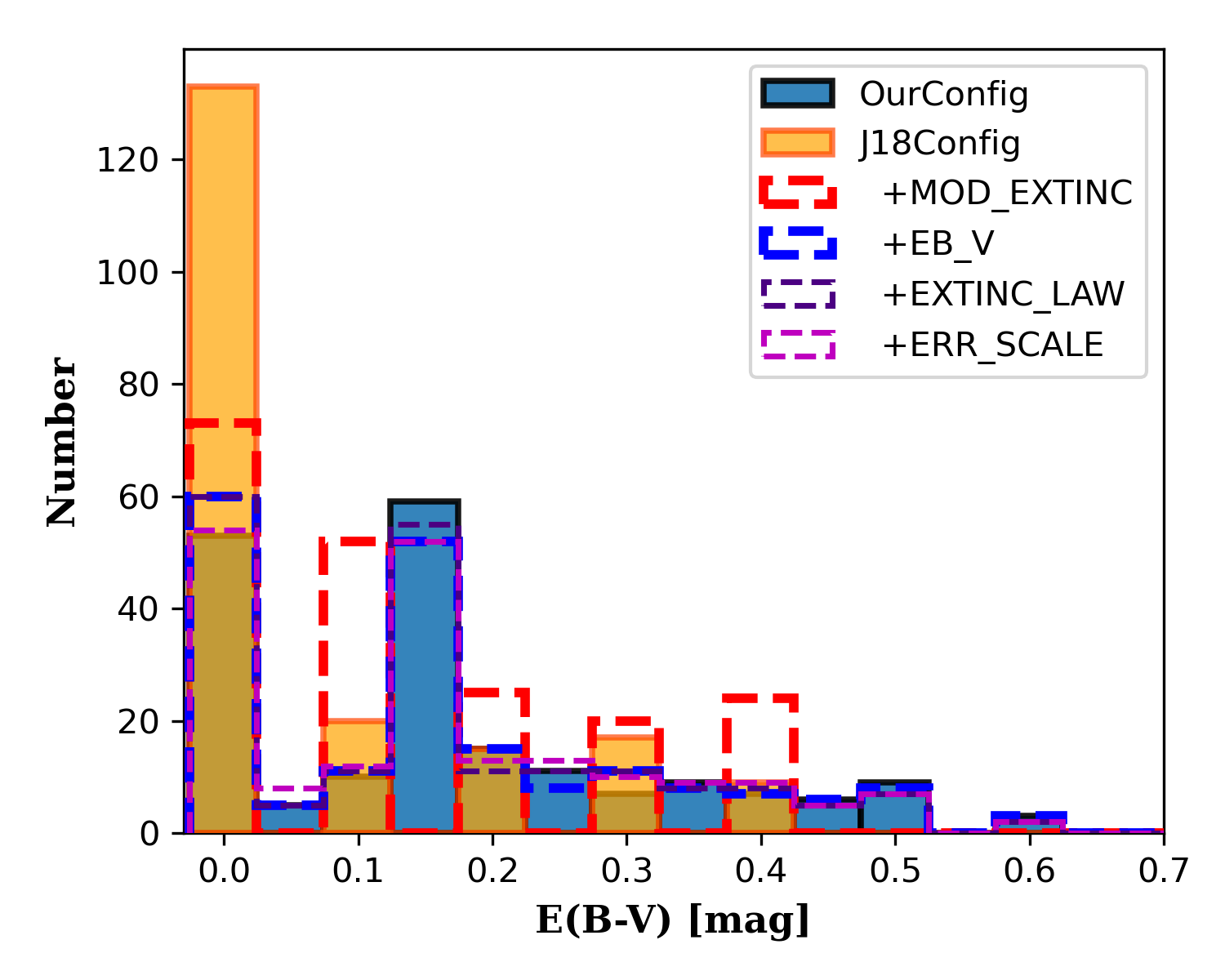}
	\caption{
	Distribution of host galaxy \ebv{} determined by \lephare{} with our (blue solid histogram), \citetalias{Jones2018} (orange solid), and \citetalias{Jones2018} + \textit{MOD\_EXTINC} (red dashed), + \textit{EB\_V} (blue dashed), + \textit{EXTINC\_LAW} (indigo dashed), and + \textit{ERR\_SCALE} (violet dashed) input parameters.
	133 out of 194 hosts have \ebv{} = 0.0 mag when using \citetalias{Jones2018} configuration, while only 53 have when using our configuration.
	Changing \textit{`MOD\_EXTINC'} parameter in the configuration file has the largest impact on the \ebv{} values (see orange solid and red dashed histograms).
	Expanding \textit{`EB\_V'} parameter, \ebv{} values spread well up to higher values.
	Adding other changed parameters show minor impacts.
	} 
	\label{fig:comp_ebv_all}
\end{figure}

\begin{figure*}
	\centering
		\includegraphics[width=0.4\textwidth]{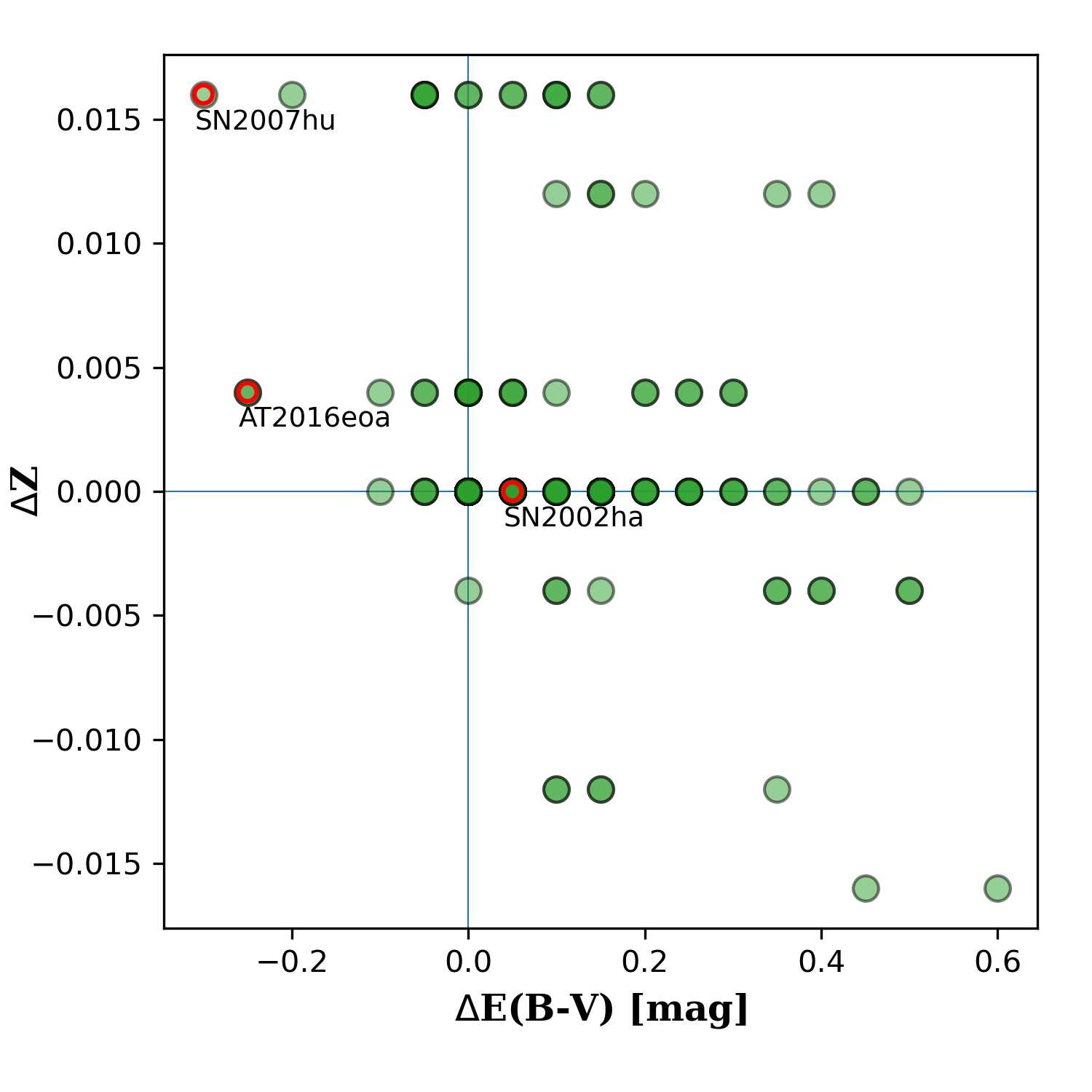}
		\includegraphics[width=0.4\textwidth]{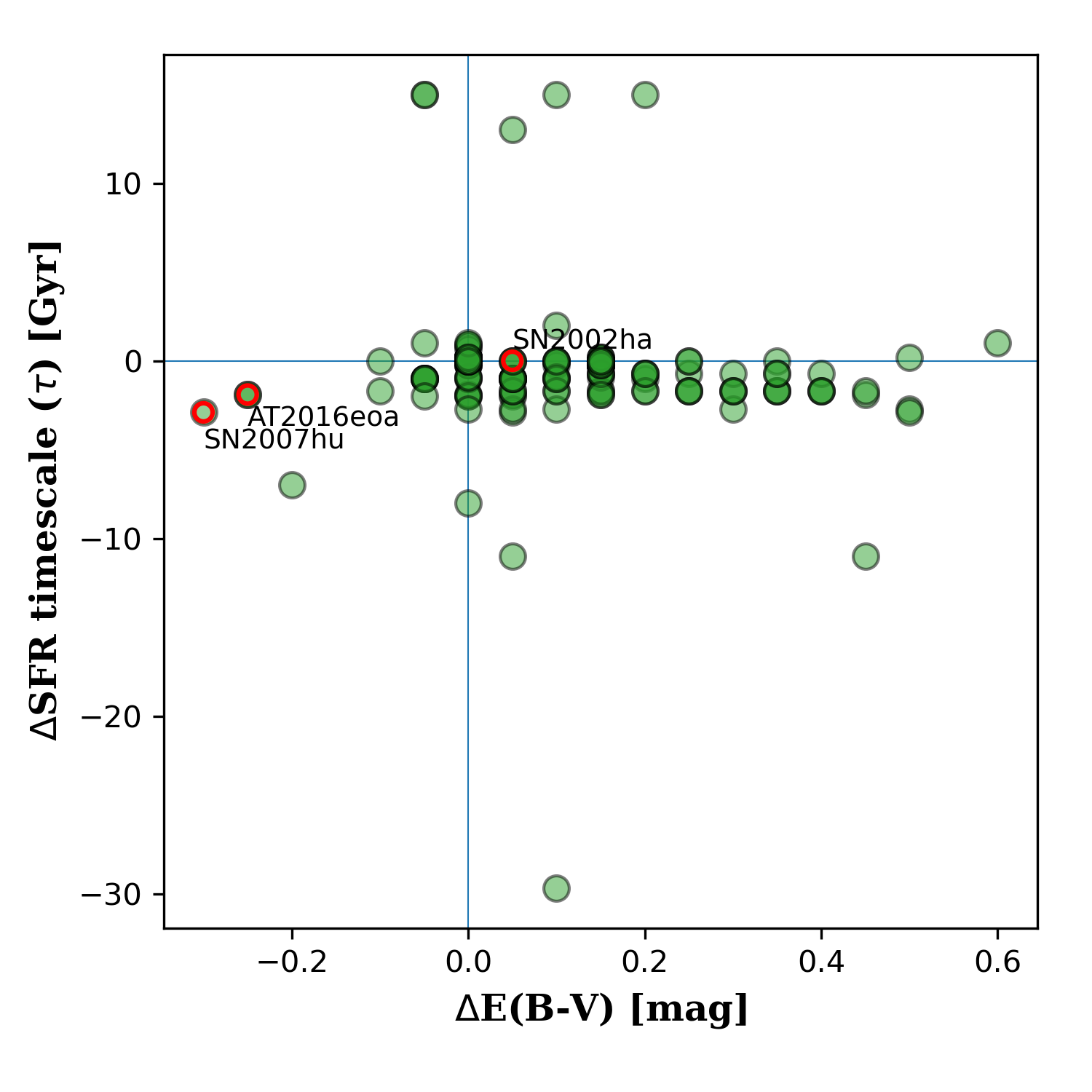}
		\includegraphics[width=0.4\textwidth]{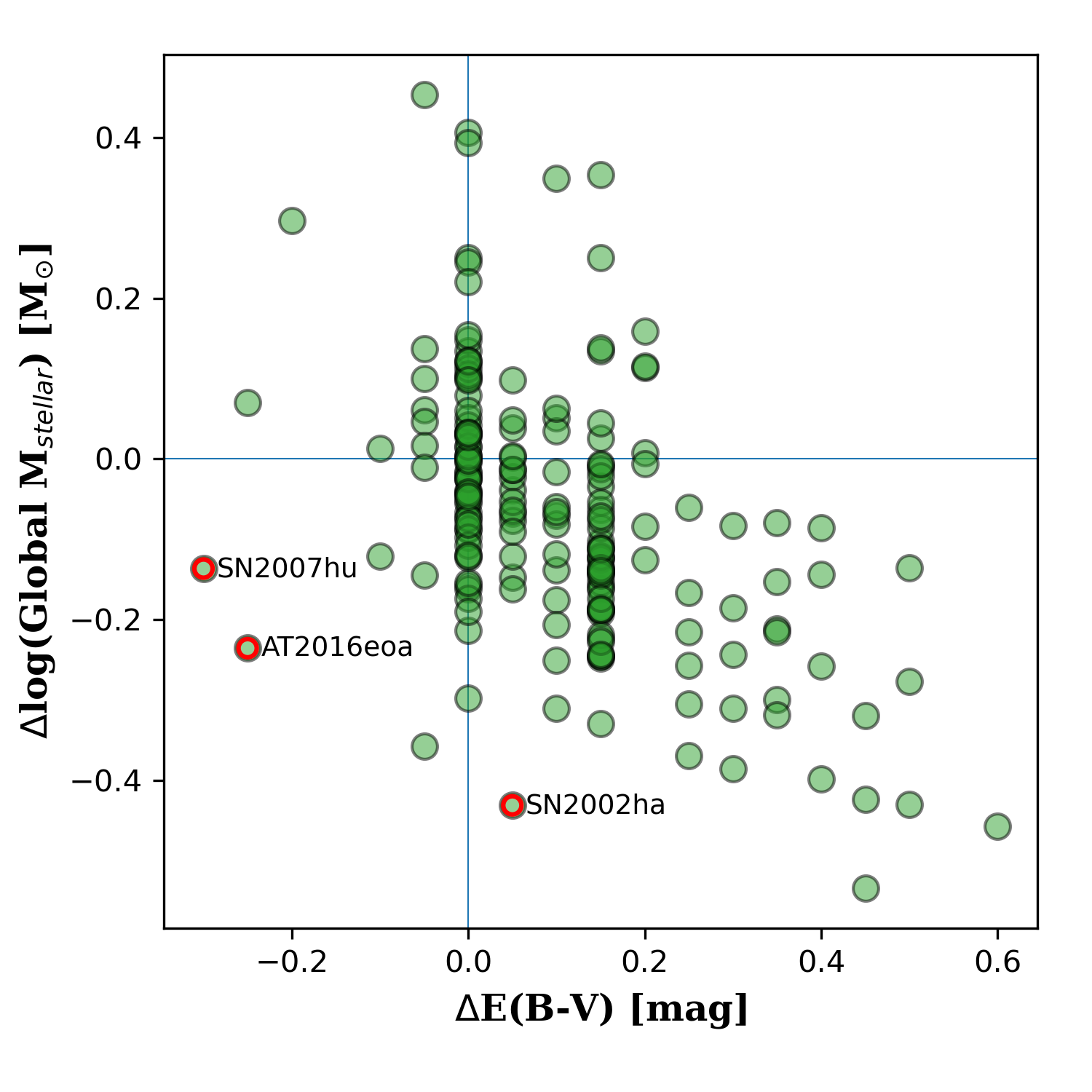}
		\includegraphics[width=0.4\textwidth]{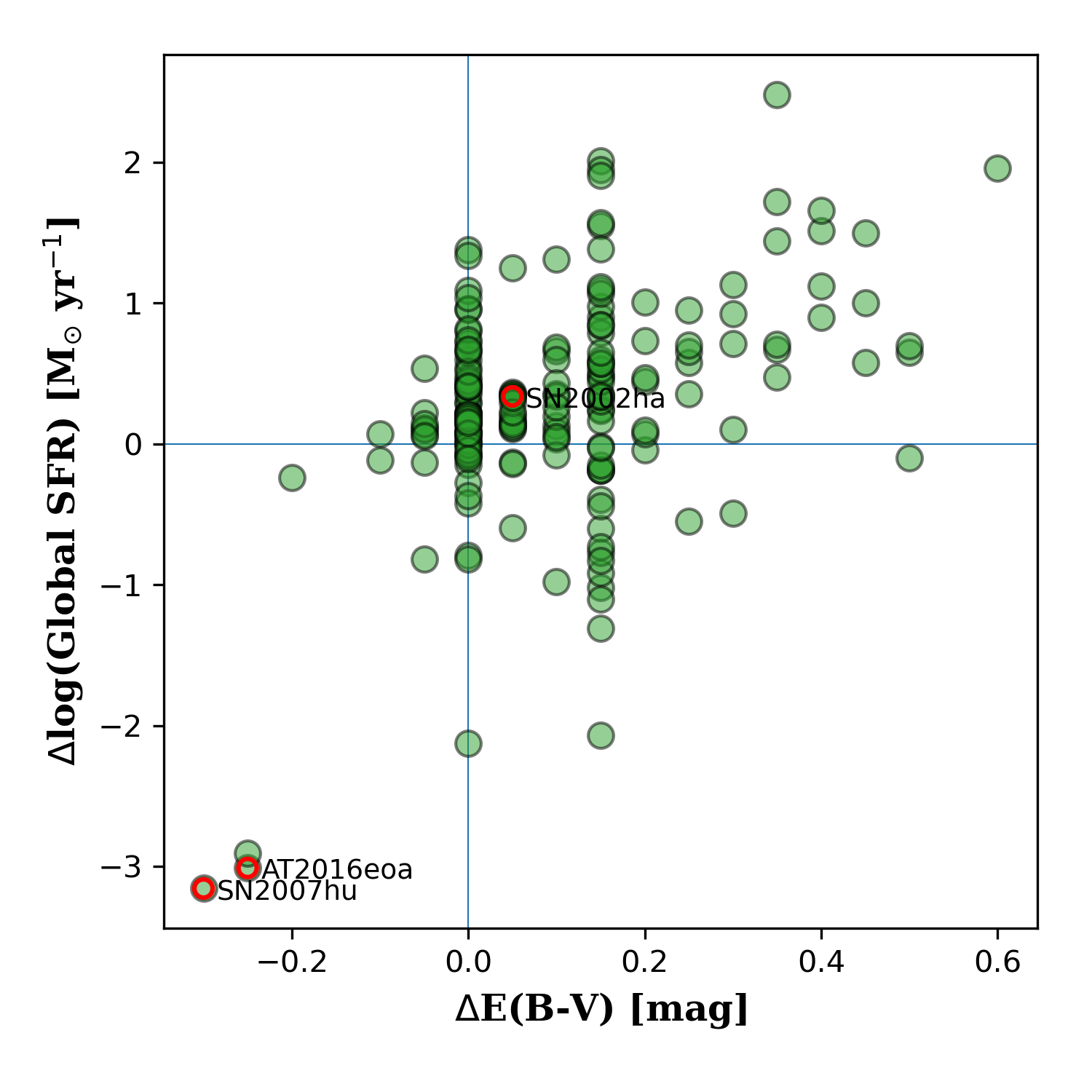}
	\caption{Differences in physical parameters as a function of \ebv{} difference determined by \lephare{} with our and \citetalias{Jones2018} configurations.
	              Three red circles indicate examples of SNe that have the largest difference in $\chi^{2}$ (see Fig.~\ref{fig:debv_dchi2}).
	              }
	\label{fig:debv}
\end{figure*}

\begin{figure}
	\centering	
	\includegraphics[width=\columnwidth]{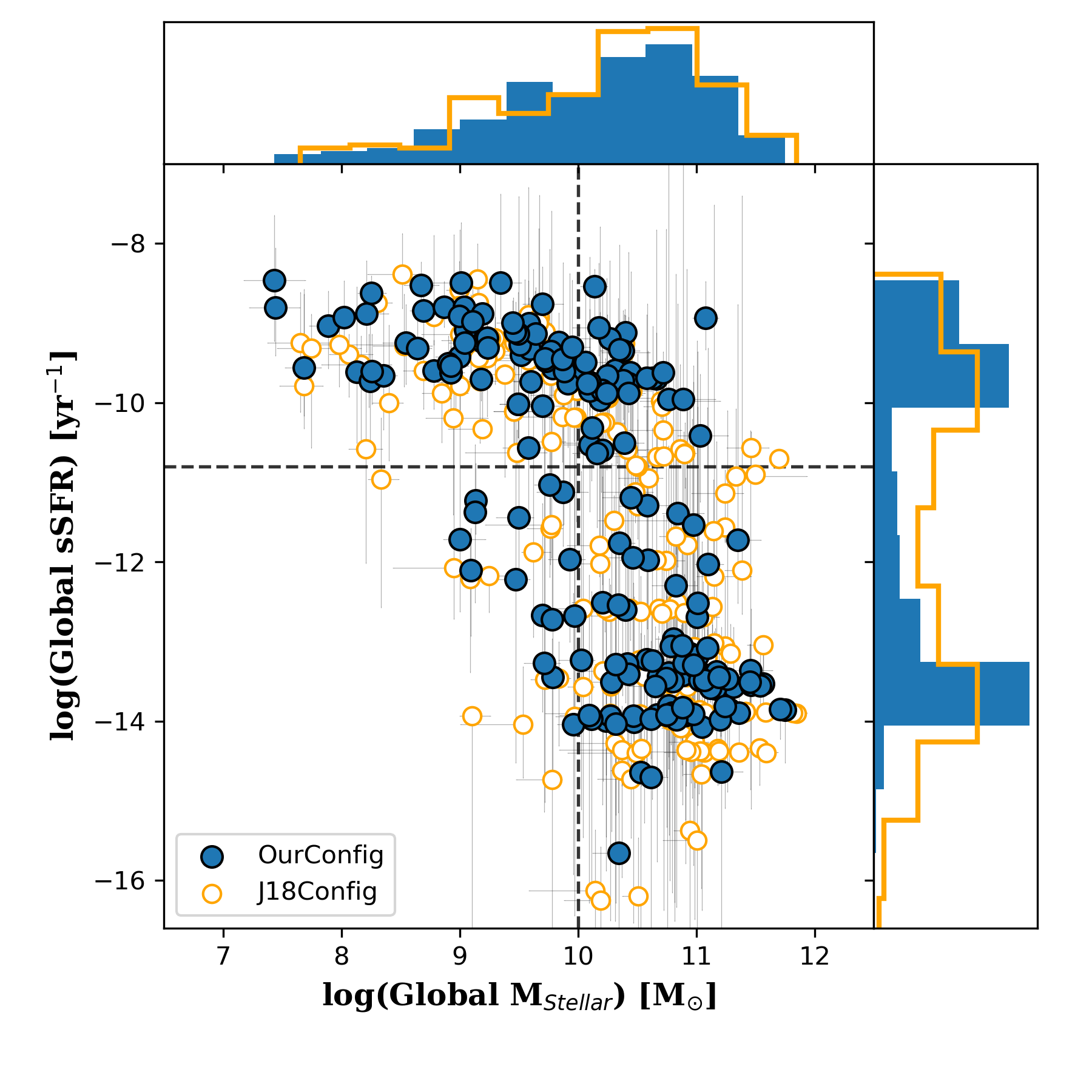}
	\caption{
	              Relation between global $M_{stellar}$ and global $sSFR$ from our (blue) and \citetalias{Jones2018} (orange) configurations.
	             Black dotted lines show our cut value to split the host sample: high-mass and low-mass galaxies are split at log($M_{\text{stellar}}$) = 10.0, and globally star-forming and passive host galaxies are at log($sSFR$) = -10.8. 
	             Using our configuration can clearly separate star-forming and passive galaxies, while using \citetalias{Jones2018} configuration can not.
	              } 
	\label{fig:mass_ssfr}
\end{figure}

\begin{figure}
	\centering	
	\includegraphics[width=\columnwidth]{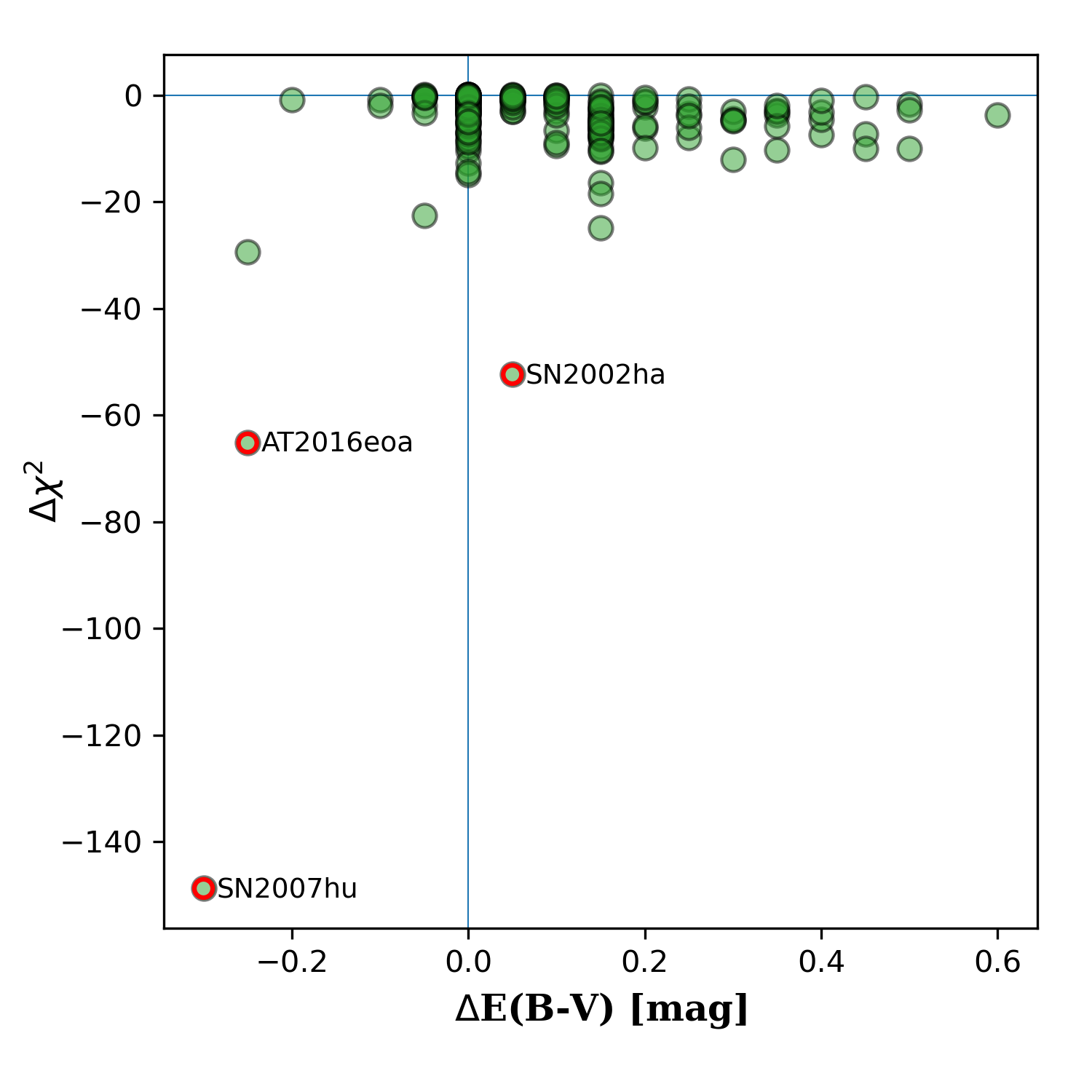}
	\caption{
	Difference in best-fitted $\chi^{2}$ returned by \lephare{} with our and \citetalias{Jones2018} configurations.
	Red circles indicate three examples with the largest $\chi^{2}$ difference.
	More negative $\Delta\chi^{2}$ value indicates a lower $\chi^{2}$ value using our configuration compared to that using \citetalias{Jones2018} configuration.
	Applying a more physically motivated treatment for the extinction, e.g., our case, shows better quality of fit.
	} 
	\label{fig:debv_dchi2}
\end{figure}

First, we check the impact of the extinction treatment on \ebv{} values.
Fig.~\ref{fig:comp_ebv_all} shows the distribution of host galaxy \ebv{} determined by \lephare{} with configurations that change step by step from \citetalias{Jones2018} to ours.
Using the \citetalias{Jones2018} configuration (orange solid histogram) results in 69\% (133 out of 194) host galaxies with \ebv{} = 0.0 mag.
Using our configuration (blue solid histogram), this fraction decreases to 27\% (53 out of 194) and peak at 0.15 mag, that is a similar to that presented in \citet{Johansson2021} (see their Fig. 9).
We find that changing the \textit{`MOD\_EXTINC'} parameter, that is the range of models for which extinction will be applied, shows the largest effect: from 69\% to 38\% host galaxies that have \ebv{} = 0.0 mag (see orange and red histograms).
As shown in the figure, when we expand the range and also add more values of the \textit{`EB\_V'} parameter, the \ebv{} values are distributed well over the entire range.

Next, we plot in Fig.~\ref{fig:debv} an impact of \ebv{} difference from different configurations on the physical parameters fitted and estimated by \lephare{}.
Different configurations return different combinations of best-fitted \ebv{} and BC03 templates that have different metallicity and SFR timescale, as shown in the upper panels.
This in turn determines the different global stellar mass and SFR for the galaxies (lower panels of Fig.~\ref{fig:debv}, and Fig~\ref{fig:mass_ssfr} with global sSFR).
The average difference in mass is 0.1 dex, and shows a similar distribution each other.
Regarding the SFR, host galaxies when using \citetalias{Jones2018} configuration has lower values of SFR: mean and median values are 0.3 dex (a factor of 2) lower than when using our configuration.
This trend is expected, such that if there is no reddening (or dust), \lephare{} considered that that galaxy is intrinsically redder, so that returns a low SFR value.
This can explain that two clear populations of star-forming and passive galaxies are observed when using our configuration, while using \citetalias{Jones2018} configuration can not, as shown in Fig.~\ref{fig:mass_ssfr}.

From the above results with the best-fitted $\chi^{2}$ values returned by \lephare{} (Fig.~\ref{fig:debv_dchi2}), we conclude that using our configuration that uses more physically motivated parameters in terms of the extinction treatment provides more robust results for the host galaxy properties.
Therefore, based on our classification, we estimate a misclassification rate when using \citetalias{Jones2018} configuration.
Among 92 star-forming (73 low mass) galaxies classified using our configuration, 15\% (8\%) are identified as passive (massive) galaxies when using \citetalias{Jones2018} configuration.
For the 102 passive (121 massive) galaxies, 3\% (1\%) are misclassified as star-forming (less massive) galaxies.

This misclassification could underestimate the true size of the environmental dependence of SN Ia luminosity \citep[see][]{Briday2022}.
This makes it difficult to find the physical origin of the dependence and to make an accurate statistical correction during the light-curve fitting process, that leads to biased cosmological parameters.
Therefore, a re-investigation of the environmental dependence in the \citetalias{Jones2018} sample is required, with more physically motivated input parameters in terms of the extinction treatment to reduce the misclassification of environments.

We note that in Fig.~\ref{fig:mass_ssfr} we present our cut criteria for high- or low-mass, globally star-forming or passive galaxies.
For the stellar mass, we divide hosts into high- and low-mass galaxies at $M_{\text{stellar}}$ = $10^{10} M_{\odot}$, which is conventionally used and well studied in many previous studies  \citep[see][]{Sullivan2010, Kim2019, Smith2020}.
For the sSFR, we take log($sSFR$) = --10.8, that can split well between star-forming and passive galaxies for our sample \citep[for similar criteria, see][]{Sullivan2010, Childress2013, Pan2014, Kim2019}.

\begin{figure*}
	\centering
	\includegraphics[width=0.49\textwidth]{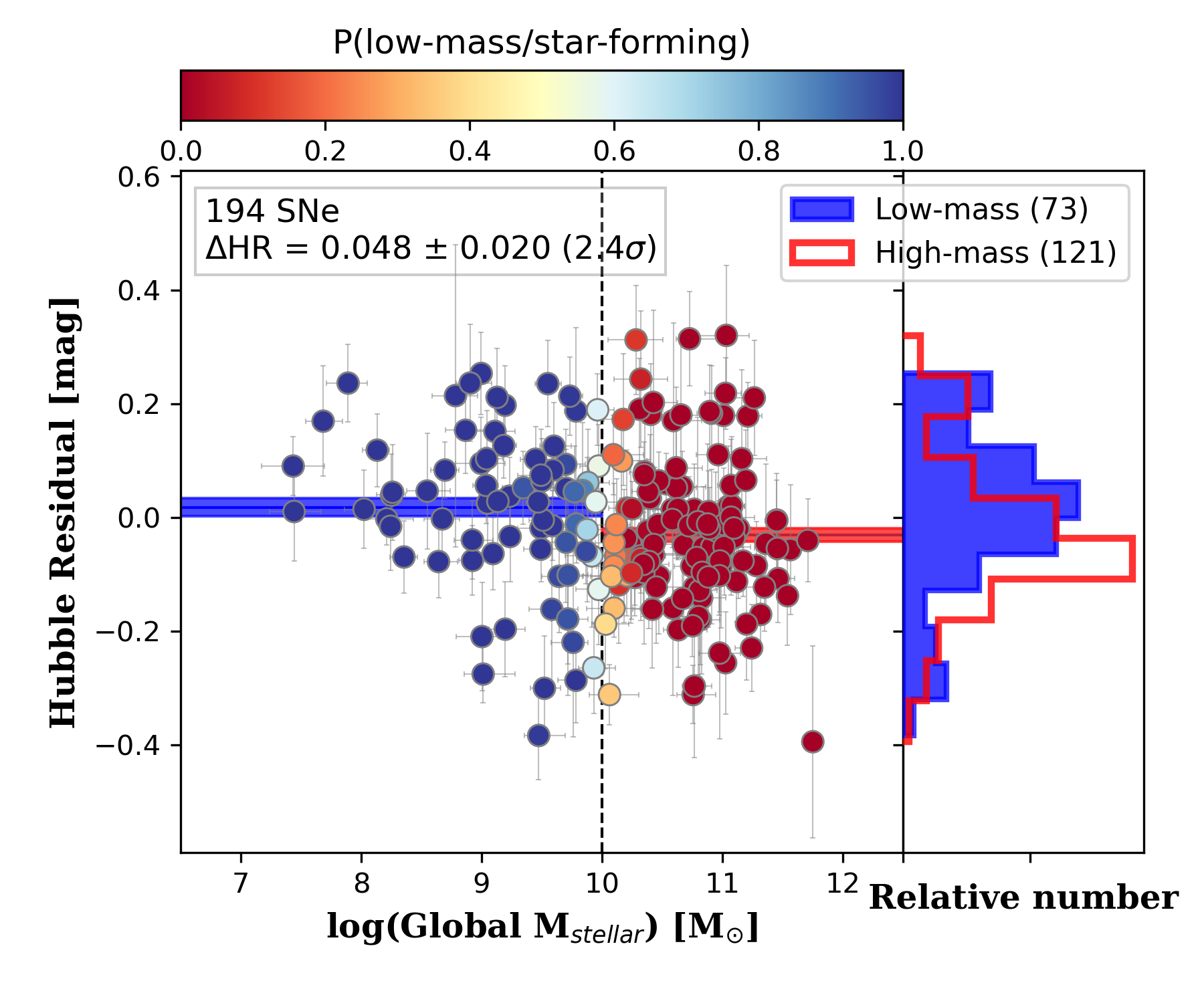}
	\includegraphics[width=0.49\textwidth]{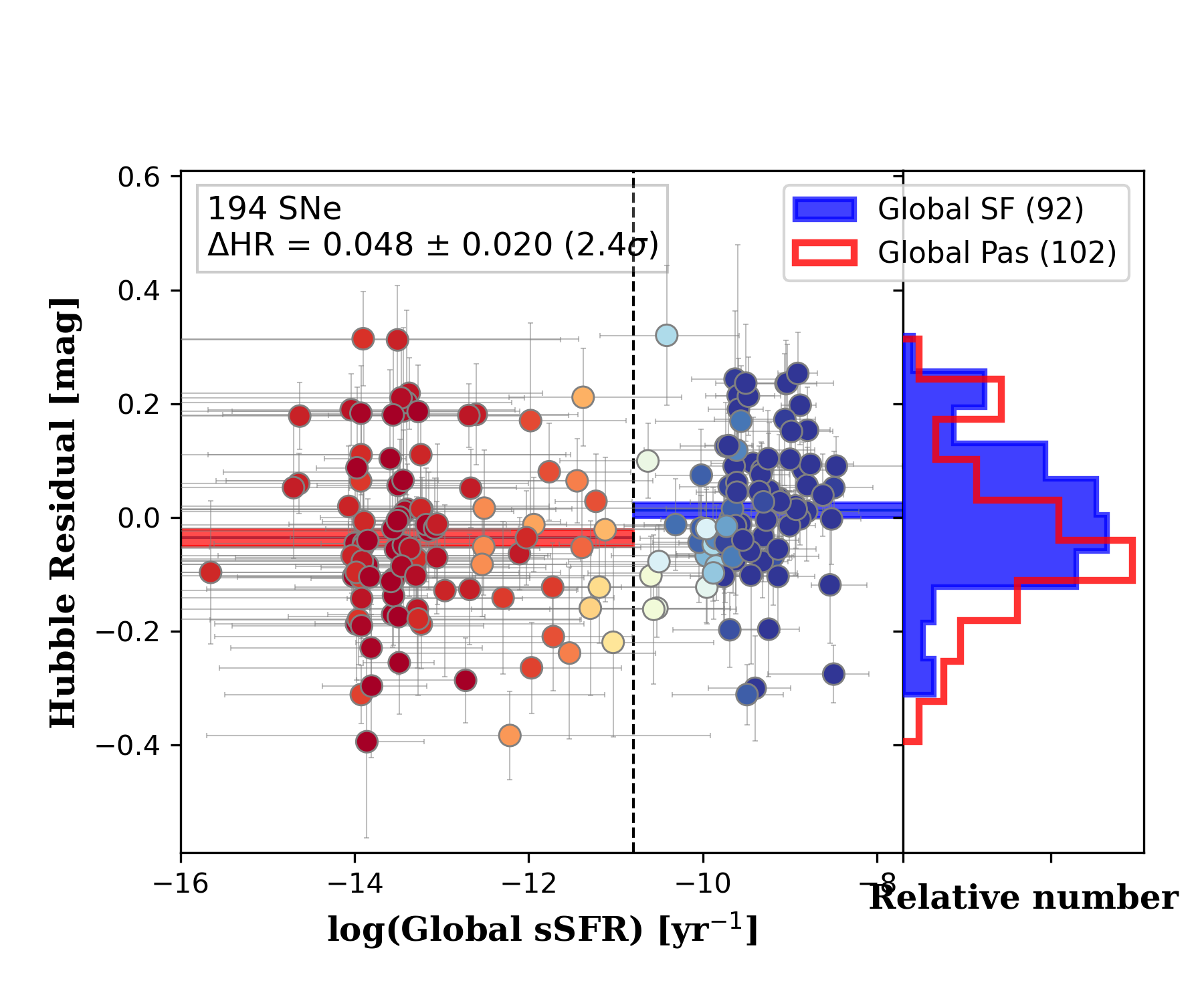}
	\caption{Dependence of SN Ia luminosities on the host global $M_{stellar}$ (left panel) and global $sSFR$ (right). 
	             SNe Ia that exploded in low-mass or star-forming host galaxies are $0.048\pm0.020$ mag ($2.4\sigma$) fainter, after applying the standard light-curve corrections plus the BBC method, than those in high-mass or passive hosts. 
	             Two horizontal bands show the weighted mean of HR in each group with the corresponding error of the mean. 
	             Vertical dotted lines present our cut value to split the host sample: high- and low-mass galaxies at log($M_{\text{stellar}}$) = $10^{10} M_{\odot}$, and star-forming and passive environments at log($sSFR$) = --10.8. 
	             The marker-color indicates the probability that the SN explodes in a low-mass (left) or star-forming host galaxy (right) (see a color-bar).}
	\label{fig:mass_gssfr_step}
\end{figure*}

\section{Reanalysing the environmental dependence of SN Ia luminosities in J18 data}
\label{sec:result}

The accuracy of local SFR measurements matters, because, as mentioned in Sec~\ref{sec:intro}, we have to access accurate photometric environmental tracers to derive precise SNe Ia distances.
We argue above that \citetalias{Jones2018}'s local SFR measurement is inaccurate.
Hence, in this section, we reanalyse the environmental dependence of SN Ia luminosities in the \citetalias{Jones2018} sample with our measurement of host galaxy properties.
We use the Hubble residuals (HRs) provided by \citetalias{Jones2018} and then estimate difference in HRs in each environment.
Their HRs are estimated based on the SALT2.4 light-curve fitter \citep{Guy2007, Guy2010, Betoule2014} with the BEAMS with Bias Corrections method \citep[BBC;][]{Kessler2017}.
To estimate the size of difference in HRs, we use the maximum likelihood approach, similar to \citetalias{Jones2018} \citep[see also][]{Jones2015, Rigault2015}.
The maximum likelihood model requires four parameters: two means and two rms scatters for two separate Gaussian distributions.
Here, the two separate Gaussian distributions correspond to high-mass and low-mass galaxies or star-forming and passive environments.
Then, the model simultaneously determines the four parameters.
From this, we calculate the difference in mean HR and its error.
The determined values for each environment are presented in Tab.~\ref{tab:step_summary}.

First, we present the dependence of SN Ia luminosities on the host global stellar mass and global sSFR in Figure~\ref{fig:mass_gssfr_step}.
SNe Ia that exploded in low-mass or star-forming environment are $0.048\pm0.020$ ($2.4\sigma$) mag fainter, after light-curve corrections, than those in high-mass or passive environment, in good agreement with many previous studies \citep[see table 9 of][for a summary of previous studies]{Kim2019}.
The size of HR difference split by host stellar mass is similar as that estimated by \citetalias{Jones2018} ($0.058\pm0.018$ mag), while they use 273 SNe Ia and their hosts with different photometry and different SED fitting model based on different templates, namely Pan-STARRS $grizy$ \citep{Chambers2016} and \texttt{Z-PEG} SED fitting code \citep{zpeg} based on P\'EGASE.2 templates \citep{Fioc1997}.

Next, we investigate the dependence of SN Ia luminosities on their local environments.
Local environments are inferred as locally passive or locally star-forming by using the method introduced by \citet{Kim2018}.
Briefly, \citet{Kim2018} collected a sample of SNe Ia from an SN Ia catalogue constructed by \citet{Kim2019} for global host properties, and  \citet{Jones2015} and \citet{Rigault2015} for local environment properties, where the local SFR was estimated from $GALEX$ far-UV data.
Then, with a sample that has both global and local properties, they showed that the SNe Ia in globally passive hosts are also in locally passive environments, while SN Ia in globally star-forming hosts are either in locally passive and locally star-forming environments.
However, when restricting SNe Ia in globally star-forming hosts to a low-mass host galaxy subset, we can only select SNe Ia in locally star-forming environments.

\begin{figure*}
	\centering
	\includegraphics[scale=0.65]{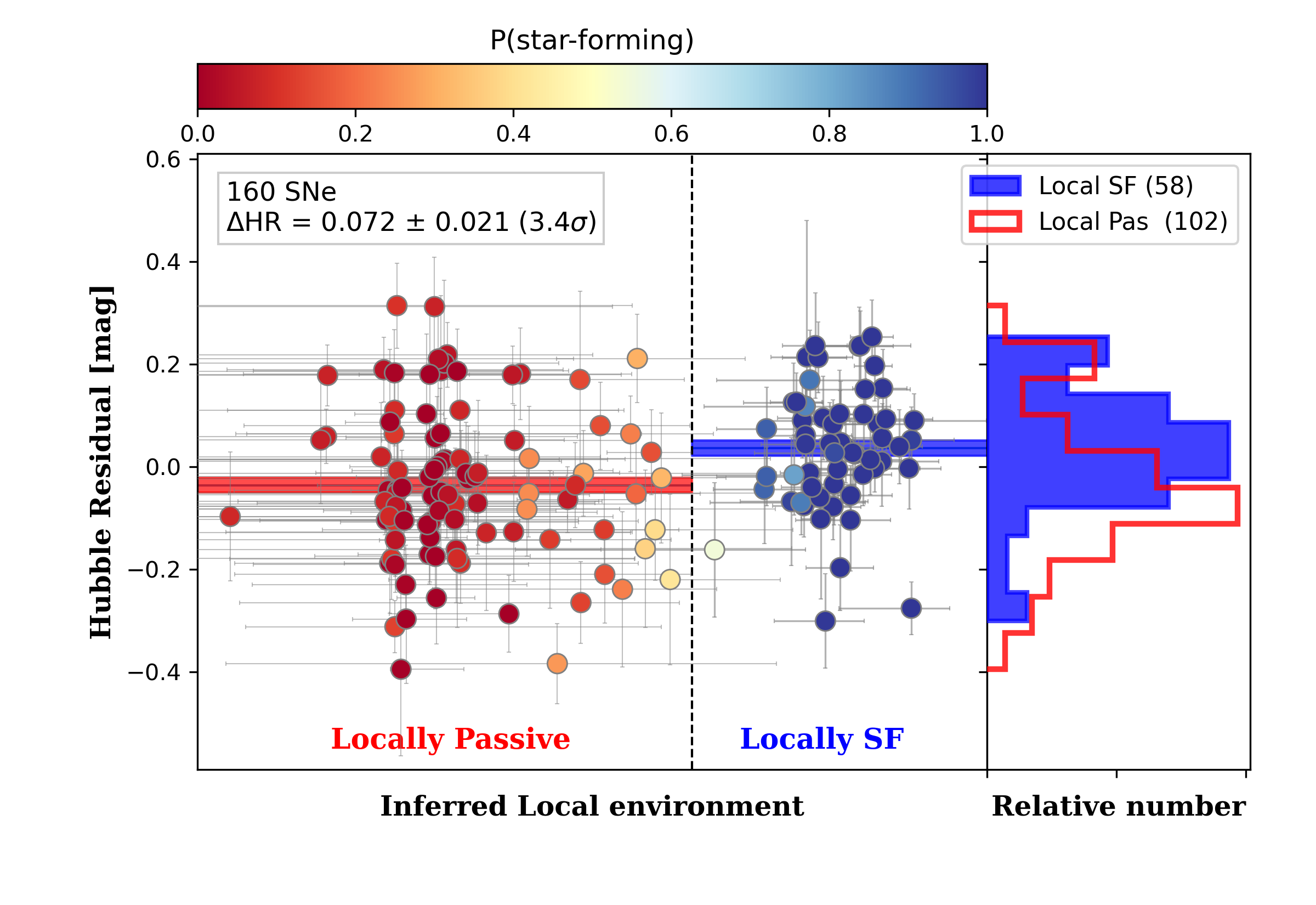}
	\caption{Same as Fig~\ref{fig:mass_gssfr_step}, but for the inferred local environment based on the method introduced by \citet{Kim2018}. 
	            We can still observe the dependence in the \citetalias{Jones2018} sample: SNe Ia in locally star-forming environments are $0.072\pm0.021$ mag ($3.4\sigma$) fainter than those in locally passive environments, after applying the standard light-curve corrections plus the BBC method. 
	            We note that the \citet{Kim2018} method only gives which local environment an SN belongs to, like locally passive or locally star-forming. 
	            Here we use global $sSFR$ values on the $x$-axis for illustrative purpose and a vertical dotted line, log($sSFR$) = --10.8, presents our cut value to split the host sample into global star-forming and passive environments.
	            }
	\label{fig:lssfr_step}
\end{figure*}

Applying this method to our sample of 194 host galaxies, split at global $M_{\text{stellar}}$ = $10^{10} M_{\odot}$ and log(global $sSFR$) = --10.8\footnote{\citet{Kim2018} split their sample at global $M_{\text{stellar}}$ = $10^{10} M_{\odot}$ and log(global $sSFR$) = --10.4. When we apply this log(global $sSFR$) split point to our sample, the sizes of global and local sSFR steps are slightly increased to $0.051\pm0.019$ and $0.074\pm0.021$, respectively.}, results in a final local sample of 160. Among the final local sample, 102 SNe Ia are in locally passive environments and 58 SNe Ia are in locally star-forming environments (Tab.~\ref{tab:step_summary})\footnote{The 34 excluded SNe Ia, which are in globally star-forming and high-mass galaxies, have a mean HRs of $-0.031\pm0.020$ mag.}.
Applying this method From this local sample, Figure~\ref{fig:lssfr_step} shows the dependence of SN Ia luminosities on their local environments.
We find that SNe Ia in locally star-forming environments are fainter than those in locally passive environments: the difference in mean HRs is $0.072\pm0.021$ mag ($3.4\sigma$; Tab.~\ref{tab:step_summary}).
Although the magnitude of the HR difference is smaller than that of \citet{Kim2018} (i.e., $0.081\pm0.018$ mag ($4.5\sigma$) with 368 SNe Ia), which employed the same method for selecting a local sample and have 23 SNe Ia in common, the environmental dependence of SN Ia luminosities is still there at a significant level in the \citetalias{Jones2018} sample.

\begin{table*}
\centering
\caption{The mean and rms scatter of SN Ia Hubble residuals in different host environments}
\label{tab:step_summary}
\begin{tabular}{l r r c}
\hline\hline\\ [-0.8em]
\multirow{2}{*}{Group} & \multirow{2}{*}{$N$} & Mean residual                                 & rms    \\ [0.1em]
                                    &                                  & \multicolumn{1}{c}{(mag)}               & (mag)  \\ [0.1em]
\hline \\ [-0.5em] 
Low-mass (log($M_{\text{stellar}}) \le 10$) & 73  & $0.017 \pm0.016$ & $0.110\pm0.014$ \\ [0.1em]
High-mass (log($M_{\text{stellar}}) > 10$) & 121 & $-0.031\pm0.012$ & $0.097 \pm0.011$ \\[0.2em]
\hline \\[-0.8em]
Diff.                                                             & 194 & $\textbf{0.048}\pm\textbf{0.020}$ \\ [0.1em]
\hline \\[-0.5em]
Globally Star-Forming                                 & 92 &  $0.012\pm0.013$ & $0.093\pm0.011$ \\ [0.1em]
Globally Passive                                         & 102 & $-0.036\pm0.015$ & $0.113\pm0.012$ \\ [0.2em]
\hline \\[-0.8em]
Diff.                                                              & 194 & $\textbf{0.048}\pm\textbf{0.020}$ \\ [0.1em]
\hline \\ [-0.5em]
Locally Star-Forming                                  & 58 &  $0.036\pm0.015$ & $0.086\pm0.014$ \\ [0.1em]
Locally Passive                                           & 102 & $-0.036\pm0.015$ & $0.113\pm0.012$ \\ [0.2em]
\hline \\ [-0.8em]
Diff.                                                              & 160 & $\textbf{0.072}\pm\textbf{0.021}$ \\ [0.1em]
\hline
\end{tabular}
\end{table*}

\section{Discussion and Conclusion}
\label{sec:discussion}

In this paper, we reconsider the photometric local SFR measurements and the argument that the dependence of SN Ia luminosities on the local SFR environment is insignificant ($0.051 \pm 0.020$ mag; $2.6\sigma$) in \citetalias{Jones2018}.
We show that their local SFR measurements estimated from $ugrizy$ photometry with \lephare{} SED fitting code are inaccurate, based on the theoretical basis of SFR measurement and the methodology they applied.
From reanalysing \citetalias{Jones2018} host sample with SDSS $ugriz$ global photometry data and our more physically motivated extinction treatments for the same \lephare{} code, we estimate global stellar mass and star formation rate.
Then, local star formation environments are inferred from them by applying the \citet{Kim2018} method.
We show that there is still local sSFR dependence of SN Ia luminosities, after empirical light-curve and further BBC corrections, with the mean HR difference of  $0.072\pm0.021$ mag ($3.4\sigma$).

The size of mean HR difference in local environment in this work is smaller than other local environment studies: 0.072 mag versus $\sim$0.100 mag.
This is expected, because:
\begin{itemize} 
\item[1)] \citetalias{Jones2018} uses the BBC method when calculating HRs. 
\citet{Scolnic2018} shows roughly $1\sigma$ larger difference when they do not implement the BBC method;
\item[2)] we estimate HR difference after SALT2 light-curve fit, not within the light-curve fit process as a third standardization parameter.
\citet{Rigault2020} shows with their sample that the HR difference is increased from 0.125 mag to 0.163 mag when fitting as a third parameter of SALT2;
\item[3)] when inferring the local environment by employing the \citet{Kim2018} method, it is expected that there will be some misclassification of the local environment.
\citet{Briday2022} shows a negative correlation between the misclassification rate (i.e., contamination) and the size of HR difference.
Therefore, the size of mean HR difference would increase if there is no contamination in classifying the environment.
\end{itemize}

The estimated step sizes in this work, therefore, should be considered as lower limits in each environment.
The true values are likely to be larger and, therefore, even more discrepant with the findings of \citetalias{Jones2018}.

The origin of this SN Ia luminosity dependence on its environment has remained elusive.
In order to estimate unbiased cosmological parameters \citep[e.g.,][]{Kim2018, Kim2019} and to solve the Hubble tension \citep[e.g.,][]{Rigault2015}, understanding the origin is important. 
We expect that this would be achieved by several thousands of SNe Ia in the various redshift range discovered by the Zwicky Transient Facility \citep{Bellm2019, Graham2019} and the Vera Rubin Observatory  Legacy Survey of Space and Time \citep{lsst2009} in the near future.

\section*{Acknowledgements}
We thank the anonymous referee for constructive suggestions to clarify the manuscript.
This project has received funding from the European Research Council (ERC) under the European Union’s Horizon 2020 research and innovation programme (grant agreement No 759194 - USNAC) and also supported by the Science and Technology Facilities Council [grant number ST/V000713/1]. 

This analysis used \textsc{\texttt{pandas}}\citep{McKinney2010}, \textsc{\texttt{numpy}}\citep{Harris2020}, \textsc{\texttt{scipy}}\citep{Virtanen2020}, and \textsc{\texttt{matplotlib}}\citep{Hunter2007}.


\section*{Data Availability}

The data underlying this article are available in the article and in its online supplementary material.
This work is based on data that are publicly available in \citetalias{Jones2018}.



\bibliographystyle{mnras}



\appendix

\section{Data of SNe Ia and their host galaxies}

\begin{table*}
\centering
\caption{Data of SNe Ia and their host galaxies used in this paper. 
	      \lephare{} values are on a log-scale.
	       $P_{low-mass}$ and $P_{SF}$ are the probability that the SN Ia explodes in a low-mass and star-forming host galaxy, respectively.
              This table is available in its entirety in machine-readable form.} 
\label{tab:data}
\begin{adjustbox}{width=1.1\textwidth,center=\textwidth}

\begin{tabular}{lcccccccccccc}
\hline\hline\\ [-0.8em]
           &                                & \multicolumn{2}{c}{\citetalias{Jones2018} }           & &\multicolumn{6}{c}{\lephare{}} \\ \cline{3-4} \cline{6-11}
     SN &     $z_{\text{cmb}}$ & HR    & Error &&  MASS\_INF & MASS\_MED & MASS\_SUP & SSFR\_INF & SSFR\_MED & SSFR\_SUP & $P_{low-mass}$ & $P_{SF}$ \\
           &                               & (mag) &         &&                     &  (M$_{\odot}$)  &                       &                   & (yr$^{-1}$)  \\
\hline \\ [-0.5em] 
        010026 & 0.03222 &     0.027 &   0.059 &&    9.800 &    9.949 &   10.189 &  -10.147 &   -9.303 &   -8.775 &   0.58 &  0.96 \\
      10028 & 0.06484 &    -0.106 &   0.067 &&   10.144 &   10.273 &   10.482 &  -15.583 &  -13.934 &  -11.365 &   0.02 &  0.11 \\
       1241 & 0.08860 &    -0.078 &   0.065 &&   10.226 &   10.392 &   10.493 &  -11.543 &  -10.504 &   -9.967 &   0.01 &  0.61 \\
      12779 & 0.07883 &     0.080 &   0.085 &&   10.227 &   10.353 &   10.502 &  -15.509 &  -11.764 &  -10.825 &   0.00 &  0.15 \\
      12781 & 0.08331 &     0.186 &   0.082 &&   10.816 &   10.895 &   10.961 &  -16.173 &  -13.271 &  -12.644 &   0.00 &  0.00 \\
      12898 & 0.08289 &    -0.005 &   0.062 &&    9.393 &    9.512 &    9.664 &   -9.601 &   -9.269 &   -9.027 &   1.00 &  1.00 \\
      12950 & 0.08163 &     0.027 &   0.061 &&    9.369 &    9.470 &    9.573 &   -9.282 &   -9.113 &   -8.925 &   1.00 &  1.00 \\
      14318 & 0.05721 &    -0.127 &   0.097 &&    9.781 &    9.973 &   10.096 &  -15.700 &  -12.677 &  -11.432 &   0.59 &  0.07 \\
      16333 & 0.07073 &    -0.188 &   0.079 &&    9.919 &   10.030 &   10.224 &  -15.610 &  -13.235 &  -11.364 &   0.39 &  0.10 \\
      16392 & 0.05834 &    -0.006 &   0.083 &&   11.358 &   11.451 &   11.515 &  -13.984 &  -13.509 &  -13.052 &   0.00 &  0.00 \\
      17240 & 0.07182 &    -0.175 &   0.102 &&   10.646 &   10.801 &   10.892 &  -14.766 &  -13.497 &  -12.503 &   0.00 &  0.00 \\
      17258 & 0.08835 &    -0.197 &   0.083 &&    9.059 &    9.196 &    9.362 &   -9.603 &   -9.242 &   -8.894 &   1.00 &  1.00 \\
      18241 & 0.09390 &     0.187 &   0.146 &&    9.687 &    9.785 &    9.867 &  -15.410 &  -13.442 &  -12.101 &   1.00 &  0.02 \\
      19899 & 0.09023 &    -0.070 &   0.063 &&    8.239 &    8.355 &    8.468 &  -10.612 &   -9.656 &   -8.978 &   1.00 &  0.88 \\
      1994M & 0.02484 &    -0.188 &   0.098 &&   11.117 &   11.201 &   11.310 &  -14.457 &  -13.980 &  -13.181 &   0.00 &  0.00 \\
      1994Q & 0.02878 &    -0.104 &   0.099 &&    9.502 &    9.643 &    9.840 &   -9.714 &   -9.136 &   -8.713 &   0.96 &  1.00 \\
      1994S & 0.01568 &    -0.064 &   0.132 &&   10.156 &   10.391 &   10.620 &  -10.308 &   -9.756 &   -9.211 &   0.05 &  0.97 \\
      19968 & 0.05578 &    -0.068 &   0.060 &&    9.968 &   10.113 &   10.235 &  -15.571 &  -13.970 &  -11.748 &   0.22 &  0.08 \\
     1996bl & 0.03424 &    -0.098 &   0.081 &&   10.065 &   10.245 &   10.512 &  -11.317 &   -9.878 &   -8.916 &   0.09 &  0.74 \\
     1998dk & 0.01209 &    -0.220 &   0.166 &&    9.577 &    9.759 &    9.885 &  -13.989 &  -11.031 &   -9.875 &   0.97 &  0.42 \\
     1998ef & 0.01663 &    -0.297 &   0.126 &&   10.664 &   10.767 &   10.913 &  -14.370 &  -13.806 &  -13.170 &   0.00 &  0.00 \\
      1999X & 0.02645 &    -0.073 &   0.109 &&   10.106 &   10.321 &   10.495 &  -16.274 &  -13.289 &  -11.402 &   0.07 &  0.09 \\
     1999cc & 0.03144 &    -0.036 &   0.083 &&   10.967 &   11.098 &   11.204 &  -14.740 &  -12.026 &  -11.122 &   0.00 &  0.09 \\
     1999dq & 0.01290 &    -0.239 &   0.151 &&   10.806 &   10.977 &   11.148 &  -13.524 &  -11.532 &  -10.542 &   0.00 &  0.23 \\
     1999ej & 0.01532 &     0.180 &   0.142 &&   10.576 &   10.656 &   10.877 &  -14.302 &  -13.558 &  -13.142 &   0.00 &  0.00 \\
     2001az & 0.04094 &     0.110 &   0.081 &&   10.781 &   10.965 &   11.094 &  -16.273 &  -13.240 &  -11.516 &   0.00 &  0.08 \\
     2001da & 0.01706 &    -0.160 &   0.153 &&   10.373 &   10.589 &   10.774 &  -13.823 &  -11.293 &   -9.752 &   0.00 &  0.37 \\
     2001en & 0.01531 &    -0.179 &   0.135 &&    9.579 &    9.715 &    9.886 &  -17.222 &  -13.270 &  -11.399 &   0.95 &  0.09 \\
     2001fe & 0.01473 &    -0.097 &   0.126 &&   10.269 &   10.348 &   10.570 &  -16.894 &  -15.655 &  -12.109 &   0.00 &  0.09 \\
     2001gb & 0.02673 &    -0.041 &   0.129 &&   10.241 &   10.441 &   10.638 &  -10.094 &   -9.620 &   -8.900 &   0.01 &  0.99 \\
     2001ic & 0.04443 &    -0.395 &   0.169 &&   11.680 &   11.751 &   11.828 &  -13.972 &  -13.860 &  -13.199 &   0.00 &  0.00 \\
      2002G & 0.03509 &    -0.142 &   0.134 &&   10.649 &   10.828 &   11.029 &  -16.206 &  -12.295 &  -11.021 &   0.00 &  0.12 \\
     2002bf & 0.02477 &    -0.129 &   0.151 &&  10.591 &   10.809 &   10.957 &  -16.221 &  -12.963 &  -11.448 &   0.00 &  0.08 \\
     2002bz & 0.03802 &     0.019 &   0.106 &&   10.881 &   11.049 &   11.183 &  -16.387 &  -14.068 &  -11.766 &   0.00 &  0.08 \\
     2002ck & 0.02988 &    -0.019 &   0.088 &&   10.970 &   11.096 &   11.204 &  -16.103 &  -13.082 &  -12.118 &   0.00 &  0.01 \\
     2002ha & 0.01268 &     0.103 &   0.156 &&   11.073 &   11.160 &   11.369 &  -14.345 &  -13.592 &  -13.411 &   0.00 &  0.00 \\
      2003U & 0.02846 &    -0.119 &   0.103 &&   10.031 &   10.141 &   10.258 &   -8.756 &   -8.539 &   -8.118 &   0.10 &  1.00 \\
     2003ae & 0.03493 &    -0.162 &   0.131 &&    9.395 &    9.583 &    9.800 &  -12.660 &  -10.564 &   -9.610 &   0.97 &  0.54 \\
     2003cq & 0.03417 &     0.020 &   0.122 &&   10.967 &   11.078 &   11.183 &   -9.396 &   -8.936 &   -8.544 &   0.00 &  1.00 \\
     2003ic & 0.05586 &    -0.384 &   0.078 &&    9.357 &    9.474 &    9.694 &  -15.699 &  -12.217 &   -9.915 &   0.99 &  0.27 \\
     2003it & 0.02532 &    -0.089 &   0.106 &&   10.860 &   10.987 &   11.098 &  -14.386 &  -13.315 &  -11.801 &   0.00 &  0.05 \\
      2004L & 0.03379 &    -0.123 &   0.099 &&   10.312 &   10.451 &   10.696 &  -14.020 &  -11.189 &   -9.740 &   0.00 &  0.39 \\
     2004as & 0.03282 &     0.211 &   0.086 &&    9.060 &    9.129 &    9.197 &  -12.224 &  -11.375 &  -10.244 &   1.00 &  0.31 \\
     2004ef & 0.03002 &    -0.083 &   0.076 &&   10.687 &   10.819 &   11.020 &  -16.370 &  -13.917 &  -11.564 &   0.00 &  0.09 \\
     2004ey & 0.01531 &    -0.022 &   0.127 &&    9.667 &    9.876 &   10.101 &  -14.001 &  -11.122 &  -10.445 &   0.71 &  0.32 \\
     2005be & 0.03411 &     0.014 &   0.080 &&   10.634 &   10.766 &   10.857 &  -16.253 &  -13.396 &  -12.196 &   0.00 &  0.02 \\
     2005hc & 0.04601 &     0.064 &   0.051 &&   10.348 &   10.471 &   10.652 &  -15.596 &  -13.931 &  -11.333 &   0.00 &  0.11 \\
     2005hf & 0.04346 &    -0.105 &   0.088 &&   10.777 &   10.884 &   10.960 &  -15.344 &  -13.827 &  -12.610 &   0.00 &  0.01 \\
     2005hj & 0.05689 &     0.074 &   0.081 &&    9.299 &    9.493 &    9.643 &  -10.541 &  -10.018 &   -9.601 &   1.00 &  0.93 \\
     2005kc & 0.01448 &    -0.000 &   0.137 &&   10.964 &   11.071 &   11.268 &  -14.397 &  -13.481 &  -13.010 &   0.00 &  0.00 \\
     2005mc & 0.02640 &    -0.013 &   0.100 &&   10.908 &   11.014 &   11.126 &  -16.299 &  -13.176 &  -12.261 &   0.00 &  0.00 \\
      2006N & 0.01402 &     0.087 &   0.142 &&   10.535 &   10.616 &   10.833 &  -14.443 &  -13.974 &  -13.482 &   0.00 &  0.00 \\
     2006ac & 0.02348 &    -0.097 &   0.098 &&   10.696 &   10.833 &   11.001 &  -16.311 &  -13.982 &  -11.631 &   0.00 &  0.09 \\
     2006al & 0.06876 &     0.110 &   0.070 &&    9.987 &   10.095 &   10.342 &  -15.684 &  -13.926 &  -11.571 &   0.19 &  0.09 \\
     2006bb & 0.02558 &    -0.020 &   0.123 &&   11.046 &   11.135 &   11.238 &  -14.331 &  -13.562 &  -13.119 &   0.00 &  0.00 \\
     2006bq & 0.02190 &    -0.024 &   0.103 &&   10.844 &   10.947 &   11.065 &  -16.296 &  -13.151 &  -12.213 &   0.00 &  0.01 \\
     2006cj & 0.06846 &     0.190 &   0.090 &&   10.224 &   10.313 &   10.505 &   -9.937 &   -9.590 &   -9.308 &   0.00 &  1.00 \\
 \hline
\end{tabular}
\end{adjustbox}
\end{table*}


\bsp	
\label{lastpage}
\end{document}